\documentclass[twocolumn,floatfix,superscriptaddress,prl]{revtex4-1}

\usepackage{times}
\usepackage{amssymb}
\usepackage{color,graphicx}
\usepackage{amsmath}
\usepackage{amsbsy}
\usepackage{amsthm}
\usepackage{bbm}
\usepackage{bm}
\usepackage{epsfig}
\usepackage{float}
\usepackage{comment}
\usepackage{soul}

\definecolor{myurlcolor}{rgb}{0,0,0.7}
\definecolor{myrefcolor}{rgb}{0.8,0,0}
\usepackage[unicode=true,pdfusetitle, bookmarks=true,bookmarksnumbered=false,bookmarksopen=false, breaklinks=false,pdfborder={0 0 0},backref=false,colorlinks=true, linkcolor=myrefcolor,citecolor=myurlcolor,urlcolor=myurlcolor]{hyperref}

\newcommand{\nn}{\nonumber\\}
\newcommand{\ket}[1]{\left| {#1} \right\rangle}
\newcommand{\bra}[1]{\left\langle {#1}\right|}
\newcommand{\braket}[2]{\langle #1|#2\rangle}
\renewcommand{\t}[1]{\textrm{#1}}
\newcommand{\sinc}{{\rm sinc }}
\newcommand{\pr}{p_{\sqcap}^\delta}  
\newcommand{\LL}{L}
\newcommand{\pd}{p_{\LL}} 
\newcommand{\ww}{\delta}
\newcommand{\bC}{C_{\chi,\tilde\varphi_\chi}}
\newcommand{\bp}{p_{\chi}}

\begin{document}
\title{$\pi$-Corrected Heisenberg Limit}
\author{Wojciech G{\'{o}}recki}
\affiliation{Faculty of Physics, University of Warsaw, Pasteura 5, 02-093 Warsaw, Poland}
\author{Rafa{\l} Demkowicz-Dobrza{\'n}ski}
\affiliation{Faculty of Physics, University of Warsaw, Pasteura 5, 02-093 Warsaw, Poland}
\author{Howard M. Wiseman}
\affiliation{Centre for Quantum Computation and Communication Technology (Australian Research Council), \\
Centre for Quantum Dynamics, Griffith University, Brisbane, Queensland 4111, Australia}
\author{Dominic W. Berry}
\affiliation{Department of Physics and Astronomy, Macquarie University, Sydney, NSW 2109, Australia}

\begin{abstract}
We consider the precision $\Delta \varphi$ with which the parameter $\varphi$, appearing in the unitary map $U_\varphi = e^{ i \varphi \Lambda}$ acting on some type of probe system, can be estimated when there is a finite amount of prior information about $\varphi$. We show that, if $U_\varphi$ acts $n$ times in total, then, asymptotically in $n$, there is a tight lower bound
$\Delta \varphi \geq \frac{\pi}{n (\lambda_+ - \lambda_-)}$,
where $\lambda_+$, $\lambda_-$ are the extreme eigenvalues of the generator $\Lambda$.
 This is greater by a factor of $\pi$ than the conventional Heisenberg limit, derived
 from the properties of the quantum Fisher
 information. That is, the conventional bound is never saturable. Our result makes no assumptions on the measurement protocol, and is relevant not only in
the noiseless case but also if noise can be eliminated using quantum error correction techniques.
 \end{abstract}
\maketitle

\emph{Introduction and statement of result.}
The \emph{Heisenberg limit} (HL) is the central concept for
the whole field of quantum metrology research
as it epitomizes
the potential of optimal quantum metrology protocols to surpass
standard schemes that are restricted by the so-called \emph{standard quantum limit} (SQL) \cite{giovannetti2006quantum,Paris2009, giovannetti2011advances,Toth2014, Demkowicz2015, Schnabel2016, degen2017quantum, Pezze2018, Pirandola2018}. For the canonical example of interferometry to measure a stationary optical phase, these two limits are expressed in terms of the number of photon passes through the unknown phase. The
HL for the estimation precision is conventionally given as $1/n$, while the SQL corresponds simply to the $1/\sqrt{n}$ shot-noise precision limit.
This scaling improvement can be achieved using entangled photon states \cite{Mitchell2004}, or multiple passes \cite{Higgins2007}, or a combination of both \cite{Daryanoosh2018}. In all such cases the essential feature is that phase is being accumulated coherently over the $n$ uses of the probe system, unlike in standard schemes where each probe (photon)
interferes only with itself and the whole procedure is repeated $n$ times gathering statistics that leads to $1/\sqrt{n}$ improvement of precision.

In a generalized phase estimation scenario, evolution of a probe system is given by a unitary $U_\varphi = \exp(i \varphi \Lambda)$, where
$\Lambda$ is an arbitrary Hermitian generator of the transformation. In what follows, we allow the spectrum of $\Lambda$ to be
arbitrary, apart from being bounded from above and from below by  $\lambda_+$ and $\lambda_-$ respectively.
Hence, the parameter $\varphi$ is not necessarily restricted to the $[0,2\pi)$ interval. Analysis of the problem using the concept of the quantum Fisher information (QFI) and the
quantum Cram{\'e}r-Rao (CR) bound leads to the following SQL and HL respectively \cite{giovannetti2006quantum}:
\begin{equation} \label{eq:conventional}
\Delta \varphi_{\t{SQL}}  \geq  \frac{1}{\sqrt{k n} (\lambda_+ - \lambda_-)}, \quad  \Delta \varphi_{\t{HL}}
\geq  \frac{1}{\sqrt{k} n (\lambda_+ - \lambda_-)},
\end{equation}
where $k $ is the number of repetitions of the experiment, $n$ is the number of applications of the unitary $U_\varphi$ in a single repetition of the experiment,  The SQL corresponds to the situation when $n$ independent interrogations of the probe system are performed in a single experiment, while the HL takes into account the most general interrogation scheme involving $n$ uses of $U_\varphi$. That may include coherent sequential probes, entangled probes, as well as the most general adaptive schemes. Interestingly, in such noiseless unitary parameter estimation scenarios there is no advantage to using adaptive strategies, as the simplest sequential scenario where the phase is being coherently imprinted on a single probe $n$ times already leads to the above stated HL.
The fundamental advantage that entanglement and adaptiveness offer emerges in the CR bound only when noise is present \cite{demkowicz2014using}.

Importantly, by the nature of the CR bound, the above
bounds are guaranteed to be saturable only in the limit of many repetitions,  $k  \rightarrow \infty$.
On the other hand, when considering $k$ repetitions, the total number of unitary operations involved 
is $kn$.
The most general way of using those resources is to allow entanglement between \emph{all}, in which case the HL in terms of total resources is $1/(kn)$ rather than $1/(\sqrt{k} n)$ scaling \cite{Hayashi2018}.
That corresponds to using $k=1$ in Eq.~\eqref{eq:conventional}, in which case saturability cannot be guaranteed.

In many papers, the saturability discussion focuses on the existence of a measurement for which the Fisher information of the corresponding probabilistic model coincides with the QFI  \cite{Braunstein1994}. Such a measurement indeed always exists, defined by the eigenbasis of the symmetric logarithmic derivative operator. However, even with such a measurement chosen, the existence of an estimator that saturates the CR bound is guaranteed in single-shot scenarios
only if the value of the parameter is known exactly beforehand 
or in the case of a narrow class of probabilistic models called the exponential family \cite{Fend59}.

In this Letter, we prove that the asymptotically tight HL includes an additional $\pi$ factor:
\begin{equation}
\label{eq:theorem}
\Delta \varphi_{\t{HL}} \geq \frac{\pi}{n (\lambda_+ - \lambda_-)}.
\end{equation}
More formally, $\lim_{n \rightarrow \infty} n \Delta \varphi_{\t{HL}} \geq \frac{\pi}{\lambda_+ - \lambda_-}$.
The condition under which we prove this bound is that any prior information is limited.
For example, one could require that there be a prior probability distribution for $\varphi$ which is bandlimited or piecewise constant, or that the measurement work over some finite region in $\varphi$-space. Most importantly, the prior information must be independent of $n$.
That is, if you are given more resources $n$, you cannot change the task by demanding a sharper prior distribution, or further restricting the region of validity of your measurement. This requirement is necessary to make the HL a meaningful concept; otherwise, the prior information is comparable to the information from the measurement itself. The $\pi$-corrected limit we derive comes precisely from eliminating the influence of any (implicit or explicit) prior.

Equation (\ref{eq:theorem}) was conjectured in \cite{Jarzyna2015}, but the argument was indirect and restricted to the standard parallel qubit phase estimation scheme with Gaussian prior, and 
the potential impact of adaptiveness was not analyzed.
Since
the HL is the key benchmark against which any theoretically conceived or experimentally implemented quantum-enhanced strategy is compared, it is essential
to phrase it as
an actual attainable limit, unlike its most commonly encountered form, Eq.~(\ref{eq:conventional}), which is not achievable even in principle.
Our result is also timely, given the recent revival of interest in quantum error correction inspired metrological protocols
which allow estimation with HL scaling even in the presence of some particular noise types \cite{sekatski2017quantum, demkowicz2017adaptive, layden2018spatial, zhou2018achieving}.
The proper phrasing of the HL is vital not only for idealized noiseless metrological scenarios  but also in the case of more realistic noisy ones.

\begin{figure*}[t!]
\includegraphics[width=0.8 \textwidth]{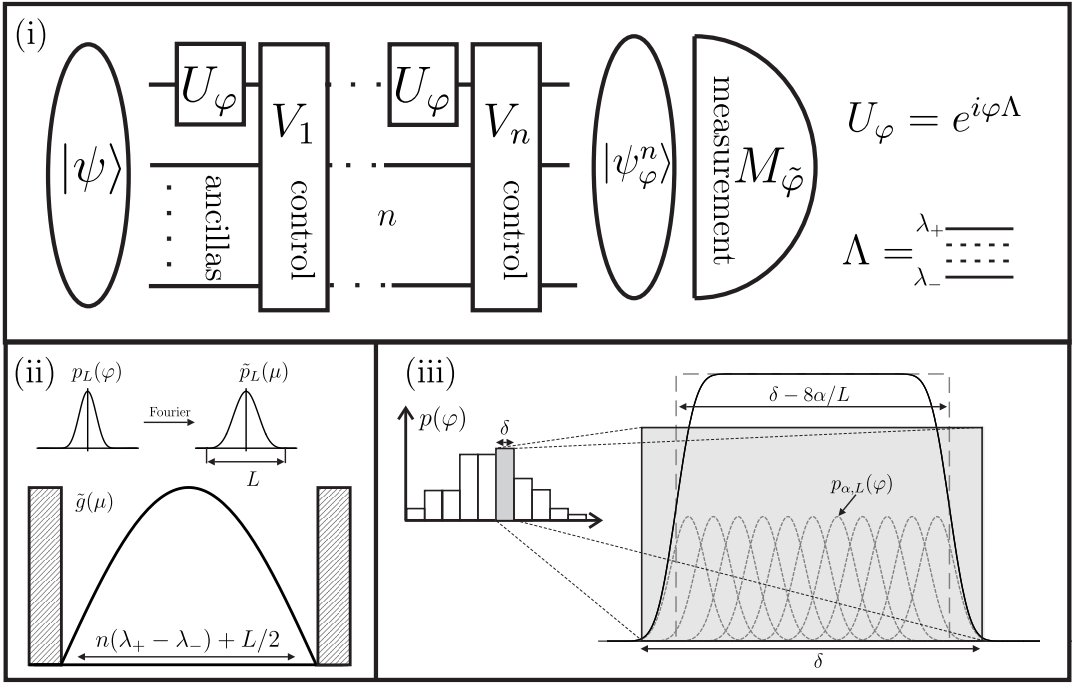}
\caption{
Graphical representation of the proof. (i) A general adaptive phase estimation protocol with the total number of phase gates $n$.
(ii) For a finite bandwidth prior $\pd(\varphi)$ we derive the bound, which is equivalent to finding the minimum energy eingenstate in an infinite potential well with width $n(\lambda_+-\lambda_-)+\LL/2$. (iii) A rectangular prior may be approximated to any desired accuracy
 by a convolution of a slightly narrower rectangular prior with a finite bandwidth Kaiser window function, to derive the bound.}
\label{fig:steps}
\end{figure*}

In our analysis, we use the Bayesian approach to estimation, also called random parameter estimation, in which a probability distribution $p(\varphi)$ is given that describes the prior knowledge of 
 $\varphi$.
In the case where no prior is known, also called non-random parameter estimation~\cite{Kay1993}, the usual approach is to require that the measurement be locally unbiased. Local unbiasedness allows one to derive the traditional (not $\pi$-corrected) form of the HL given in Eq.~\eqref{eq:conventional}, and
ensures that, typically, the estimator approximately achieves the claimed mean-square error (MSE) over some
finite range of $\varphi$.
However, it does not ensure that this finite region is independent of $n$. Since the true value of $\varphi$ cannot be known before making the measurement, a useful measurement must work over a fixed region,
even as $n$ increases.
Thus it is appropriate to consider the average of the MSE over some fixed region.
Mathematically, that is equivalent to Bayesian estimation with a flat prior over the region, and so
there is no loss of generality in using a Bayesian approach.

It is necessary 
 to exclude pathological priors,  
which could lead to arbitrarily high precision.
We do this by requiring the prior to be well approximated by a finite bandwidth function (i.e.\ a function whose Fourier transform has bounded support). 
As we will show, this is not a very restrictive condition; in particular, the results will be valid for any prior that may be approximated by a weighted sum of flat priors with nonzero fixed width $\ww >0$: $p(\varphi) \approx  \sum_{l = -\infty}^{\infty} p(l \ww)\, \Theta\!\left(\tfrac{1}{2}\ww - |\varphi - l\ww|\right)$, where $\Theta$ is the Heaviside step function.
That includes the case where the MSE is averaged over a finite region, used to address cases with an unknown prior. 
Provided the above regularity condition holds, the actual form of the prior becomes irrelevant for sufficiently large $n$. The intuition behind this result is that, in the limit of $n \rightarrow \infty$, the amount of data that can potentially be gathered on the parameter $\varphi$ is unlimited and hence overwhelms any impact of the prior on the final precision.

\emph{Proof of result.} Consider the most general adaptive estimation scheme, Fig.~\ref{fig:steps} (i).
Here $\ket{\psi}$ is the input state of a probe system potentially entangled with an arbitrary number of ancillary systems, and $V_i, i\in\{1,\dots,n\}$ are control unitary operations applied between interrogation steps where the unknown parameter is imprinted on the probe system. The final state at the output $\ket{\psi_\varphi^n}$
is measured using a generalized measurement described by a positive operator valued measure ${\{M_{\tilde{\varphi}}\t{d}\tilde{\varphi}\}}$,
where the index $\tilde{\varphi}$ represents the estimated value of the parameter upon attaining that outcome from the measurement.
The minimal expected (in the Bayesian sense) mean-square error in the estimation thus reads:
\begin{align}
&\Delta^2 \varphi =  \nonumber \\
& \min_{\ket{\psi}, \{M_{\tilde{\varphi}}\}, \{V_i\}}
\iint \t{d}\tilde\varphi\, \t{d}\varphi\,
p(\varphi) \bra{\psi_\varphi^n} M_{\tilde{\varphi}} \ket{\psi^n_{\varphi}} (\tilde{\varphi} - \varphi)^2. \label{eq:shouldbeq}
\end{align}

Let us analyze the structure of the state $\ket{\psi}$ as it evolves through the subsequent gates and control operations.
Each gate multiplies components of the state, as decomposed in the $\Lambda$ eigenbasis, by one of the $e^{i \varphi \lambda}$ factors ($\lambda$ represents some eigenvalue of $\Lambda$), while control operations $V_i$ perform a basis change. In the end, after $n$ coherent interrogations of the unknown parameter $\varphi$, the final state will
have the following structure:
\begin{equation} \label{eq:outputgeneral}
\ket{\psi_\varphi^n} = \int_{n\lambda_-}^{n\lambda_+} c(\mu) e^{i \varphi \mu} \ket{g_\mu} \t{d}\mu,
\end{equation}
where $c(\mu)$ are complex amplitudes and $\ket{g_\mu}$ are some normalized vectors which in general will not be orthogonal.
The key feature of this state is that it has bandwidth bounded by $n(\lambda_+-\lambda_-)$, as will any inner products with this state.

Now let $\pd(\varphi)$ be the prior with a finite bandwidth $\LL$, so that its Fourier transform is supported on an interval of length $\LL$.
It is known that in all single-generator unitary estimation problems with quadratic cost one may restrict the class of measurements to rank-one projective ones \cite{helstrom1976quantum, Macieszczak2014}.
Let us assume for a moment a fixed input state $\ket{\psi}$ and a measurement basis $\{\ket{\chi}\}$ with a corresponding estimator $\tilde\varphi_\chi$.
The corresponding cost reads
\begin{equation}
\label{baycost}
\Delta^2\varphi=\int \t{d}\chi \int\t{d}\varphi \, \pd(\varphi) |\braket{\psi^n_\varphi}{\chi}|^2(\tilde{\varphi}_\chi-\varphi)^2.
\end{equation}
Now, from~\cite{boas1945inequalities},
it is possible to find a function $w_\LL(\varphi)$ with bandwidth $\LL/2$ such that $|w_\LL(\varphi)|^2=\pd(\varphi)$.
Let us define functions $f_\chi(\varphi)=\braket{\psi^n_\varphi}{\chi}$ (which have bandwidths bounded by $n(\lambda_+-\lambda_-)$) and $g_{\chi}(\varphi)=w_\LL(\varphi) f_\chi(\varphi)$. We have:
\begin{equation}
\Delta^2\varphi=\int \t{d}\chi \int\t{d}\varphi \, |g_\chi(\varphi)|^2(\tilde{\varphi}_\chi-\varphi)^2.
\end{equation}
The product of two bandlimited functions gives a function which is
bandlimited by the sum of the bandwidths, so $g_\chi(\varphi)$ has a bandwidth at most $n(\lambda_+-\lambda_-)+\LL/2$.
We can then write
\begin{equation}
\label{pcb}
\Delta^2\varphi =\int \bp \bC  \t{d}\chi  \geq \min_{\chi}\bC,
\end{equation}
with
\begin{align}
\label{pxi}
\bp&=\int |g_\chi(\varphi)|^2\t{d}\varphi,\\
\bC &=\frac{1}{\bp}\int |g_\chi(\varphi)|^2(\tilde\varphi_\chi-\varphi)^2  \t{d}\varphi .
\end{align}
The task of minimizing $\bC$ for a given $\tilde\varphi_\chi$ is equivalent to that of minimisation for $\tilde\varphi_\chi=0$, because the optimal $g_\chi(\varphi)$ can be shifted by any amount without altering the bandwidth. We may also set $\bp=1$ (this just sets the normalization of the function $g$). After applying the Fourier transform, the minimization problem reads
\begin{equation}
\min_{\tilde g(\mu)}\int_{n\lambda_--\LL/4}^{n\lambda_++\LL/4} \left|\frac{\t{d} \tilde g(\mu)}{\t{d} \mu}\right|^2 \t{d}\mu,
\end{equation}
subject to the constraints
\begin{equation}
\begin{split}
&\int_{n\lambda_--\LL/4}^{n\lambda_++\LL/4} {|\tilde g(\mu)|^2}\,  \t{d}\mu=1,\\
&\tilde g(n\lambda_--\LL/4)=0,\quad \tilde g(n\lambda_++\LL/4)=0.
\end{split}
\end{equation}
Here the boundary conditions are due to the restricted bandwidth of $g(\varphi)$.
This is equivalent to the problem of finding the minimum energy eigenstate in an infinite potential well with width $n(\lambda_+-\lambda_-)+\LL/2$. The solution for $\tilde g(\mu)$ that achieves the minimum is a sine curve (see Fig.~\ref{fig:steps}(ii)), which gives
$\min \bC = {\pi^2}/{[n(\lambda_+-\lambda_-)+\LL/2]^2}$, and hence
\begin{equation}
\label{eq:theoremnadwidth}
\Delta^2\varphi \ge \frac{\pi^2}{[n(\lambda_+ - \lambda_-)+\LL/2]^2}.
\end{equation}
That is, provided the prior has a limit $\LL$ on its bandwidth, the Heisenberg limit \eqref{eq:theorem} holds in the limit of
large $n$---i.e.\ when the prior correction $\LL/2$ becomes negligible.

Now we need only prove that any reasonable prior $p(\varphi)$ may be well approximated by a finite bandwidth function.
To first show this informally, let us introduce a family of non-negative normalized finite bandwidth functions
\begin{equation}\label{eq:paL}
p_{\alpha,\LL}(\varphi)=\mathcal{N}_{\alpha}
\LL \,{\rm sinc}^4\left(\pi \alpha\sqrt{(L\varphi/4 \alpha)^2-1}\right),
\end{equation}
where $\LL$ is the bandwidth, $\alpha$ is a parameter that controls the size of the tails and $\mathcal{N}_\alpha$ is a normalization factor that $\approx{4\sqrt{2}\pi^4\alpha^{7/2}}{e^{-4\pi\alpha}}$ for $\alpha$ large.
This $p_{\alpha,\LL}(\varphi)$ is proportional to the fourth power of the Fourier transform of the Kaiser window \cite{Kaiser} of width $\LL/4$, so has bandwidth $\LL$.
The function $p_{\alpha,\LL}(\varphi)$ has width $8\alpha/L$, beyond which
the tails are exponentially suppressed in $\alpha$, like $e^{-4\pi\alpha}$. Thus for
large $\LL$ it approximates the Dirac delta function and therefore any reasonable prior
may be approximated by its convolution with $p_{\alpha,\LL}$, i.e.\ $p(\varphi)\approx (p_{\alpha,\LL}\ast p)(\varphi)$,
for which the bandwidth is also $\LL$.

In particular, we may show that, for any prior of the form $p(\varphi) \approx  \sum_{l = -\infty}^{\infty} p(l \ww)\, \Theta\!\left(\tfrac{1}{2}\ww - |\varphi - l\ww|\right)$,
  the following bound holds:
 \begin{equation}\label{exbound}
 \Delta^2 \varphi \ge \frac{\pi^2}{[n(\lambda_+-\lambda_-)]^2}\left[ 1-  \sqrt{\frac{8\log\Delta}{\Delta}}\right],
\end{equation}
where $\Delta = n(\lambda_+-\lambda_-)\delta$.
This gives
\begin{equation}
\lim_{n \rightarrow \infty} n^2 \Delta^2 \varphi_{{\rm HL}} \geq
\frac{\pi^2}{(\lambda_+ - \lambda_-)^2},
\end{equation}
which after taking the square root of both sides proves Eq.~\eqref{eq:theorem}.

Here we just sketch the reasoning leading to the above claim, whereas the complete proof with all the technical details is
presented in the Supplementary Materials~\cite{SM}.
First, we lower bound the minimal cost in the case that the  prior is actually a single rectangular prior of width $\ww$ at $l=0$. Note that this will also be a legitimate lower bound for the original problem where the prior is a weighted sum of such rectangular priors, as the optimal strategy for this original problem cannot perform better than the optimal strategy when we additionally know to which $\ww$ interval the value of our parameter is restricted. Then, we approximate the rectangular prior via a distribution obtained by convolving a
slightly narrower rectangular distribution (narrower by $8\alpha/L$) with the $p_{\alpha,\LL}$ function, as defined in Eq.~\eqref{eq:paL}---see Fig.~\ref{fig:steps}(iii).
By setting $\alpha$ and $L$ appropriately, we can guarantee that all the deviations of the resulting cost due to modifications of the prior from the strictly rectangular one
introduce no more than $\widetilde{\mathcal O}\left(\Delta^{-1/2}\right)$ (
$\widetilde{\mathcal O}$ indicates 
that logarithmic multipliers are ignored) 
relative correction compared with the cost corresponding
 to the $p_{\alpha,\LL}$ distribution. Finally, the cost corresponding to the $p_{\alpha,\LL}$ distribution
  can be bounded using Eq.~\eqref{eq:theoremnadwidth} thanks to the finite bandwidth property of $p_{\alpha,\LL}$.

In the case of non-random parameter estimation \cite{Kay1993},
we 
average the
MSE over 
a $\varphi$-interval of size $\delta$.
This is 
equivalent to using a flat prior of width $\delta$, 
so our bound holds in this case also, and this is true regardless of whether or not the estimators are unbiased.   
The conventional scenario of unbiased estimators with no average does typically mean 
that the measurement works well
for $\varphi$ within some finite region.
However, the size of that region may depend on $n$.  
That is, as $n$ increases the region of validity for an unbiased measurement may shrink with $n$. This is exactly the case for the NOON states \cite{NOON} that maximize the CR bound, saturating the conventional HL in Eq.~(\ref{eq:conventional}). For these states  the probability distributions obtained are periodic in $\varphi$ with period $\pi/n$,
and the interval in $\varphi$ where the CR-saturating measurement is useful shrinks exactly as fast in $n$ as the CR bound itself.

To reiterate, to make the HL a meaningful concept, the bound should not rely on using more and more prior information as $n$ increases, because that would mean that the prior information 
can be comparable to the information from the measurement itself.
To derive a HL that describes the information obtained from the measurement, one must do as we have done, and take a prior that is independent of $n$, or
require a region of validity that is independent of $n$.
Then the information about $\varphi$
in the limit $n\to\infty$ comes only from the measurement on the state. Eliminating the influence of
prior information is what gives our additional factor of $\pi$.

\emph{Discussion.}
Having derived the bound let us now discuss its saturability.
It is known that in the case of a standard phase estimation problem with $u_\varphi =e^{i\varphi \sigma_z/2}$ applied in parallel to
$n$ qubits the optimal Bayesian strategy for flat prior distribution $p(\varphi) = \frac{1}{2\pi}$ ($\varphi \in [-\pi,\pi]$)
in the limit of large $n$ yields $\Delta \varphi \to \pi/n$ \cite{Berry2000}. Note also, that the optimal strategy involves application of the so called covariant measurements \cite{Holevo1982}, and a covariant measurement strategy will yield the same average cost irrespective of the form of prior. Hence the $\pi/n$ limit is saturable in the case of an arbitrary prior as well. In the case of an estimation problem with a general generator $\Lambda$, we can say that provided the prior is supported on an interval smaller than $2\pi/(\lambda_+ - \lambda_-)$
we can directly adapt the reasoning from the standard qubit phase estimation scheme by considering our elementary system as a qubit with only two accessible states being the eigenstates of $\Lambda$ corresponding to $\lambda_+$ and $\lambda_-$. This way we obtain $\Delta \varphi_{{\rm HL}} = \pi/[n(\lambda_+ - \lambda_-)]$.
However, if our prior is broader, then
clearly using this strategy we will not be able to discriminate between phases that differ by a multiple of $2\pi/(\lambda_+ - \lambda_-)$
as they effectively would lead to the same output state. In order to discriminate between these phases we would need to use additional eigenstates of $\Lambda$ corresponding to intermediate eigenvalues $\lambda$ (provided they are available).
If we use levels corresponding to eigenvalues that differ by $\epsilon$, we may discriminate between all phases which differ by less than $2 \pi / \epsilon$. Note that for our purposes, the minimal level splitting $\epsilon$ may be effectively obtained as a difference between sums of certain finite number of energy levels $\epsilon = \sum_{i \in \{i_1,\dots,i_s\}} \lambda_i - \sum_{j \in \{j_1,\dots,j_s\}} \lambda_j$, and a result may be smaller than the minimal level splitting in the $\Lambda$ itself.
Since the discrimination error drops exponentially with the number of resources used, we may sacrifice sublinear (in $n$) uses of the channel  for the purpose of this additional discrimination task and this will not affect the final scaling.

The phase estimation problem may be viewed as a special case of a more general frequency estimation problem, where
the probe system is allowed to be interrogated for the total interrogation time $T$ and the goal is to estimate a frequency-like
parameter $\omega$ entering into the Hamiltonian of the system as $H = \omega G$, with $G$ being some Hermitian operator.
The total interrogation time $T$ may be split into a number of shorter evolution steps each lasting time $t = T/n$.
Assuming the prior distribution $p(\omega)$ satisfies the regularity assumption and can be written as a sum of
rectangular priors of some finite width $\delta_\omega$ we may repeat the whole reasoning as presented above
by formally identifying $\varphi = \omega t$, $\Lambda = -G$, $n = T/t$ and arrive at
\begin{equation}
\Delta \omega \geq \frac{\pi}{T(\lambda_+ - \lambda_-)}.
\end{equation}
This is the valid asymptotically saturable bound for the most general frequency estimation adaptive strategies in the limit of long total interrogation time $T$. In particular in all the cases where, despite presence of noise, the Heisenberg scaling is being recovered via e.g.\ application of quantum error-correction inspired techniques \cite{kessler2014quantum, dur2014improved, arrad2014increasing, demkowicz2017adaptive, sekatski2017quantum, zhou2018achieving, layden2018spatial}, it is the above bound that should be used as a operationally meaningful figure of merit of such protocols and not the standard QFI based one.

Finally, since Eq.~\eqref{exbound} is not tight for finite $n$, the form of the exact achievable bound in nonasymptotic case is an interesting open question for future research.

\begin{acknowledgments}
We thank Michael Hall, Marcin Jarzyna, Pawe{\l} Kasprzak, Jan Kolodynski and Yuval Sanders for fruitful discussions. WG and RDD acknowledge support from the National Science Center (Poland) grant No.\ 2016/22/E/ST2/00559.
DWB is funded by Australian Research Council Discovery Projects DP160102426 and DP190102633.
HMW is funded by the Australian Research Council Centre of Excellence Program CE170100012.
\end{acknowledgments}

\onecolumngrid
\newpage
\appendix
\begin{center}
{\large \bf{Supplemental Material}}
\end{center}

\renewcommand{\thesection}{\Alph{section}}
\twocolumngrid
\onecolumngrid

\section{Lower bounding the Bayesian cost using rectangular priors}
Given the minimal Bayesian cost as in Eq.~(3) with a prior that is a weighted sum of rectangular priors
\begin{equation}
 p(\varphi) =  \sum_{l = -\infty}^{\infty} p(l \ww)\, \Theta\!\left(\tfrac{1}{2}\ww - |\varphi - l\ww|\right),
\end{equation}
we can obtain the following lower bound:
\begin{align}
\Delta^2 \varphi &=   \min_{\ket{\psi}, \{M_{\tilde{\varphi}}\}, \{V_i\}}
\iint \t{d}\tilde\varphi\, \t{d}\varphi\,
\sum_{l=-\infty}^\infty p(l \ww)\, \Theta\!\left(\tfrac{1}{2}\ww - |\varphi - l\ww|\right) \bra{\psi_\varphi^n} M_{\tilde{\varphi}} \ket{\psi^n_{\varphi}} (\tilde{\varphi} - \varphi)^2 \nonumber  \\
& \geq \sum_{l=-\infty}^\infty p(l \ww) \min_{\ket{\psi^{(l)}}, \{M^{(l)}_{\tilde{\varphi}}\}, \{V^{(l)}_i\}}
\int_{(l-1/2)\delta}^{(l+1/2)\delta}\t{d}\varphi\, \int \t{d}\tilde\varphi\,
  \bra{\psi^{(l) n}_\varphi} M^{(l)}_{\tilde{\varphi}} \ket{\psi^{(l)n}_{\varphi}} (\tilde{\varphi} - \varphi)^2 \nonumber \\
  & =
  \min_{\ket{\psi}, \{M_{\tilde{\varphi}}\}, \{V_i\}}
\frac{1}{\delta}\int_{-\delta/2}^{\delta/2}\t{d}\varphi\, \int \t{d}\tilde\varphi\,
  \bra{\psi^{ n}_\varphi} M_{\tilde{\varphi}} \ket{\psi^{n}_{\varphi}} (\tilde{\varphi} - \varphi)^2.
\end{align}
In the first line we have used Eq.\ (3) from the main text and substituted the expression for $p(\varphi)$.
In the second line, the inequality is due to the fact that we can optimize the estimation strategy for each rectangular prior separately, and we have also limited the integral over $\varphi$ to the region where $\Theta$ is nonzero.
In the third line, we have used the fact that the cost corresponding to shifted rectangular priors should be the same.
That is, given the optimal strategy for a given rectangular prior, we can obtain an equally good estimation strategy for a prior shifted by amount $l\delta$,
by first counter-rotating the input state by a phase $-l\delta$, performing
the same measurement as before and adding the appropriate shift $l \delta$ into the finally estimated value.
The factor of $1/\delta$ arises because the normalization of $p(\varphi)$ means that the sum over $p(l\delta)$ must be $1/\delta$.
This implies that the minimal Bayesian cost for the prior which is a weighted sum of rectangular priors of width $\delta$ cannot be smaller than the minimal Bayesian cost for a problem with a prior which is a single normalized rectangular prior of width $\delta$.

Note that this same approach works for any prior that is obtained from a convolution
\begin{equation}
p(\varphi) = \int \t{d} \theta \, p_0(\theta) q(\varphi-\theta)\, .
\end{equation}
Following the same reasoning as above,
\begin{align}
\Delta^2 \varphi &=   \min_{\ket{\psi}, \{M_{\tilde{\varphi}}\}, \{V_i\}}
\iint \t{d}\tilde\varphi\, \t{d}\varphi\,
\int \t{d} \theta \, p_0(\theta) q(\varphi-\theta)
 \bra{\psi_\varphi^n} M_{\tilde{\varphi}} \ket{\psi^n_{\varphi}} (\tilde{\varphi} - \varphi)^2 \nonumber  \\
& \geq \int \t{d} \theta \, p_0(\theta) \min_{\ket{\psi^{(\theta)}}, \{M^{(\theta)}_{\tilde{\varphi}}\}, \{V^{(\theta)}_i\}}
\iint \t{d}\varphi\, \t{d}\tilde\varphi\, q(\varphi-\theta)
  \bra{\psi^{(\theta) n}_\varphi} M^{(\theta)}_{\tilde{\varphi}} \ket{\psi^{(\theta)n}_{\varphi}} (\tilde{\varphi} - \varphi)^2 \nonumber \\
  & =
\min_{\ket{\psi}, \{M_{\tilde{\varphi}}\}, \{V_i\}}
\iint \t{d}\varphi\,\t{d}\tilde\varphi\,  q(\varphi)
  \bra{\psi^{n}_\varphi} M_{\tilde{\varphi}} \ket{\psi^{n}_{\varphi}} (\tilde{\varphi} - \varphi)^2 .
\end{align}
That is, if the prior probability distribution is obtained by convolving with $q$, then the mean-square error can be lower bounded by that using $q$ as the prior.
For the case of rectangular priors above, $q$ would be a rectangle function of width $\delta$, and $p_0$ would be a series of delta functions with height $p(l\delta)$ spaced apart by $\delta$.

\section{Normalization for finite-bandwidth function $p_{\alpha,\LL}(\varphi)$}
Here we discuss some features of the family of non-negative normalized finite bandwidth
functions:
\begin{align}
p_{\alpha,\LL}(\varphi)&=
\mathcal{N}_{\alpha} \LL \,{\rm sinc}^4\left(\pi \alpha\sqrt{(L\varphi/4 \alpha)^2-1}\right) \nn
&=\mathcal{N}_{\alpha} \frac{\LL \,{\rm sinh}^4\left(\pi \alpha\sqrt{1-(L\varphi/4 \alpha)^2}\right)}{ \left(\pi \alpha\sqrt{1 -(L\varphi/4\alpha)^2}\right)^4},
\end{align}
where $\mathcal N_\alpha$ is a normalization constant depending on $\alpha$. As mentioned in the main text, this is the fourth power of the Fourier transform of the Kaiser window function of bandwidth $L/4$, and hence has a bandwidth $L$.

First note that the function significantly changes its behavior at $\varphi=4\alpha/L$.
For $\varphi<4\alpha/L$ the sinh is exponentially large in $\alpha$, which must be compensated for by the normalization factor $\mathcal N_\alpha$ being exponentially small.
In contrast, for $|\varphi|>4\alpha/L$ the function becomes the fourth power of a sinc, so decreases as $\sim \varphi^{-4}$.
Moreover, the exponentially small normalization factor means that these tails are exponentially small in $\alpha$.
Now we would like to estimate the value of the normalization factor $\mathcal N_\alpha$ for large values of $\alpha$. When integrating the function within the interval  $[-4\alpha/L,+4\alpha/L]$ we can obtain the following approximation
\begin{align}
\int_{-4\alpha/L}^{4\alpha/L} \t{d}\varphi \, \frac{\LL \,{\sinh}^4\left(\pi \alpha\sqrt{1-(L\varphi/4 \alpha)^2}\right)}{ \left(\pi \alpha\sqrt{1 -(L\varphi/4\alpha)^2}\right)^4} &=
\frac {4\alpha}{L}\int_{-1}^{1} \t{d}x \, \frac{L\, \sinh^4\left(\pi\alpha\sqrt{1-x^2}\right)}{\pi^4 \alpha^4 (1-x^2)^2}  \nn
&= \frac {8}{\alpha^3 \pi^4}\int_{0}^{1} \t{d}x \, \frac{\sinh^4\left(\pi\alpha\sqrt{1-x^2}\right)}{(1-x^2)^2} \nn
&= \frac {8}{\alpha^3 \pi^4}\int_{0}^{1} \t{d}z \, \frac{\sinh^4\left(\pi\alpha z\right)}{z^3\sqrt{1-z^2}} \nn
&\approx \frac {1}{2 \alpha^3 \pi^4}\int_{0}^{1} \t{d}z \, \frac{e^{4\pi\alpha z}+e^{-4\pi\alpha z}}{\sqrt{1-z^2}} \label{app1}\\
&= \frac {1}{2 \alpha^3 \pi^4}\int_{-1}^{1} \t{d}z \, \frac{e^{4\pi\alpha z}}{\sqrt{1-z^2}}\nn
&= \frac {1}{2 \alpha^3 \pi^4}\pi I_0(4\pi\alpha)\nn
&\approx \frac {1}{2 \alpha^3\pi^3}\frac{e^{4\pi\alpha}}{\sqrt{8\pi^2\alpha}} \label{app2} \\
&= \frac {e^{4\pi\alpha}}{4\sqrt{2} \pi^4\alpha^{7/2}}.
\label{norm}
\end{align}
In \eqref{app1} we note that the hyperbolic sine is dominated by $z\sim 1$, so we can replace the $z^3$ by 1 in the denominator. We also omitted the terms $\propto e^{\pm 2\pi\alpha z}$ and $\propto 1$ in $\sinh^4(\pi\alpha z)$ as they have much smaller impact than $e^{4\pi\alpha z}$.
We have kept $e^{-4\pi\alpha z}$ because it enables the simplification in the next line.
In \eqref{app2} we have approximated the zeroth-order modified Bessel function of the first kind by $I_0(x)\approx {e^x}/{\sqrt{2\pi x}}$. Now we can bound the impact of sinc tails as
\begin{align}
2\int_{4\alpha/L}^{+\infty} \t{d}\varphi\, L\,\sinc^4\left(\pi \alpha\sqrt{(L\varphi/4\alpha)^2-1}\right) &=
\frac{8\alpha}{L}\int_{1}^{+\infty} \t{d}x \, L\,\sinc^4\left(\pi \alpha\sqrt{x^2-1}\right) \nn
&=8\alpha\int_{1}^{2}\t{d}x\,\, \sinc^4\left(\pi \alpha\sqrt{x^2-1}\right)+
8\alpha\int_{2}^{+\infty}\t{d}x\,\, \sinc^4\left(\pi \alpha\sqrt{x^2-1}\right) \nn
&\leq
8\alpha\int_{1}^{2}\t{d}x+
8\alpha\int_{2}^{+\infty}\t{d}x\, \left(\pi \alpha\sqrt{x^2-1}\right)^{-4}\nn
&=8\alpha\left(1+\frac{1/3-\log(3)/4}{\pi^4\alpha^4}\right).
\end{align}
We see that this quantity, when compared with \eqref{norm}, is negligible for large $\alpha$ and therefore $\mathcal N_\alpha\approx  {4\sqrt{2} \pi^4\alpha^{7/2}}{e^{-4\pi\alpha}}$.

\begin{figure*}[t!]
\includegraphics[width=0.5 \textwidth]{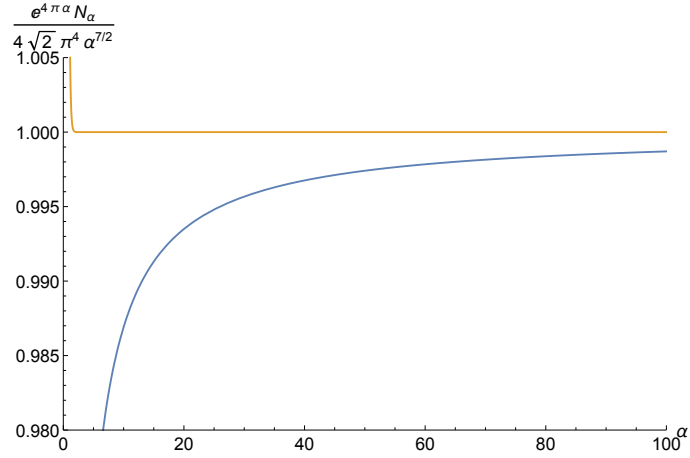}
\caption{\label{Nalpha}The value of $\mathcal N_\alpha$ is shown as a ratio to ${4\sqrt{2} \pi^4\alpha^{7/2}}{e^{-4\pi\alpha}}$ (in blue), showing that the approximation is quite accurate and $\mathcal N_\alpha<{4\sqrt{2} \pi^4\alpha^{7/2}}{e^{-4\pi\alpha}}$.
The curve in orange is the ratio of $\mathcal N_\alpha$ to the expression in Eq.~\eqref{Nseries}, showing that it is highly accurate.}
\end{figure*}

By expanding $1/z^3$ in a Taylor series about $1$ and using the asymptotic expansion of modified Bessel functions, it is also possible to obtain corrections as
\begin{equation}\label{Nseries}
\mathcal N_\alpha\approx  {4\sqrt{2} \pi^4\alpha^{7/2}}{e^{-4\pi\alpha}}\left( 1 - \frac{13}{32 \pi \alpha} - \frac{319}{2^{11} \pi^2 \alpha^2}
- \frac{10007}{2^{16} \pi^3 \alpha^3}
- \frac{1793365}{2^{23} \pi^4 \alpha^4}
- \frac{99317267}{2^{28} \pi^5 \alpha^5}
- \frac{12817002203}{2^{34} \pi^6 \alpha^6}
- \mathcal{O}(\alpha^{-7}) \right).
\end{equation}
Note that all correction terms are negative.
In practice we find that $\mathcal N_\alpha< {4\sqrt{2} \pi^4\alpha^{7/2}}{e^{-4\pi\alpha}}$, as well as it being an accurate approximation, as shown in Figure \ref{Nalpha}.

\section{Proof of the bound for the estimation cost in case of a rectangular prior in the limit of large $n$}

Now we will demonstrate how the result valid for a finite-bandwidth $p_L(\varphi)$ prior may be utilized to get a bound for the rectangular prior of size $\delta$, which will be denoted as $\pr(\varphi)$.
The most intuitive way to approximate the prior $\pr(\varphi)$ by a finite bandwidth function is to convolve it with $p_{\alpha,\LL}$. However, the distribution constructed in such a way would differ from the initial one by some non-negligible parts that stick outside of the original rectangular distribution. To avoid this problem, we consider a slightly narrowed rectangular prior of width $\delta-8{\alpha}/{L}$ and only then perform the convolution
\begin{equation}
\label{smeared}
p_{\alpha,L,\delta}(\varphi)=\frac{1}{\delta-8\alpha/L}\int_{-\delta/2+4\alpha/L}^{\delta/2-4\alpha/L}\t{d} \eta  \, p_{\alpha,L}(\varphi-\eta).
\end{equation}
This procedure guarantees that only the sinc-like tails will stick outside the range of the initial rectangular distribution after the convolution is performed.

Note that
\begin{align}
\int_{-\delta/2+4\alpha/L}^{\delta/2-4\alpha/L}\t{d} \eta  \, p_{\alpha,L}(\varphi-\eta) <\int_{-\infty}^{\infty}\t{d} \eta  \, p_{\alpha,L}(\varphi-\eta) =1 .
\end{align}
For $\varphi\in [-\delta/2,+\delta/2]$ we have $\pr(\varphi)=1/\delta$, so
\begin{equation}
\pr(\varphi)\geq \left(1-\frac{8\alpha}{\delta L}\right)p_{\alpha,L,\delta}(\varphi).
\end{equation}
Let $M^{\t{opt}}_{\tilde{\varphi}}$ represent the optimal measurement/estimator strategy for the rectangular prior $\pr(\varphi)$.
Then $M^{\t{opt}}_{\tilde{\varphi}}$ must be $0$ for  $\tilde{\varphi} \notin [-\delta/2,\delta/2]$ as the optimal strategy never estimates a value of the parameter outside of the support of the prior.
Therefore we obtain
 \begin{equation}
 \Delta^2 \varphi_{\delta} = \int\limits_{-\infty}^{+\infty} \t{d}\varphi \, \pr(\varphi)\int\limits_{-\delta/2}^{\delta/2} \t{d}\tilde{\varphi} \, \braket{\psi_\varphi^n}{M^\t{opt}_{\tilde\varphi}|\psi_\varphi^n} (\tilde\varphi-\varphi)^2 \geq
  \left(1-\frac{8\alpha}{\delta L}\right)\int\limits_{-\delta/2}^{\delta/2} \t{d}\varphi \, p_{\alpha,L,\delta}(\varphi)\int\limits_{-\delta/2}^{\delta/2} \t{d}\tilde{\varphi} \, \braket{\psi_\varphi^n}{M^\t{opt}_{\tilde\varphi}|\psi_\varphi^n}(\tilde{\varphi}-\varphi)^2.
 \end{equation}
If we  extend the limits of integration over $\varphi$ on the right side of the inequality to $(-\infty,+\infty)$, we may lower bound the resulting quantity by the minimal conditional cost corresponding to the finite bandwidth prior $p_{\alpha,L,\delta}(\varphi)$, i.e.\ ${\pi^2}/{(n(\lambda_+-\lambda_-)+L/2)^2}$. Therefore, we have
\begin{equation}
\label{boundcomp}
\Delta^2\varphi_\delta\geq \left(1-\frac{8\alpha}{\delta L}\right)\frac{\pi^2}{(n(\lambda_+-\lambda_-)+L/2)^2}-\mathcal R_{\alpha,L,\delta}\, ,
\end{equation}
where $\mathcal R_{\alpha,L,\delta}$ corresponds to the tails of the distribution
\begin{align}
\mathcal R_{\alpha,L,\delta} &=
\left(1-\frac{8\alpha}{\delta L}\right)\int_{-\infty}^{-\delta/2}\t{d}\varphi \, p_{\alpha,L,\delta}(\varphi) \int_{-\delta/2}^{\delta/2} \t{d}\tilde{\varphi} \,\braket{\psi_\varphi^n}{M^\t{opt}_{\tilde\varphi}|\psi_\varphi^n}(\tilde\varphi-\varphi)^2 \nn
& \quad +
\left(1-\frac{8\alpha}{\delta L}\right)\int_{+\delta/2}^{+\infty}\t{d}\varphi \, p_{\alpha,L,\delta}(\varphi) \int_{-\delta/2}^{\delta/2} \t{d}\tilde{\varphi} \,\braket{\psi_\varphi^n}{M^\t{opt}_{\tilde\varphi}|\psi_\varphi^n}(\tilde\varphi-\varphi)^2.
\end{align}
To bound $\mathcal R_{\alpha,L,\delta}$ we use the following chain of inequalities (we start with the right tail):
\begin{align}
&\left(1-\frac{8\alpha}{\delta L}\right)\int_{\delta/2}^{\infty}\t{d}\varphi \, p_{\alpha,L,\delta}(\varphi) \int_{-\delta/2}^{\delta/2} \t{d}\tilde{\varphi} \, \braket{\psi_\varphi^n}{M^\t{opt}_{\tilde\varphi}|\psi_\varphi^n}(\tilde\varphi-\varphi)^2\nn
&\leq\left(1-\frac{8\alpha}{\delta L}\right)\int_{\delta/2}^{\infty}\t{d}\varphi \, p_{\alpha,L,\delta}(\varphi)(\varphi+\delta/2)^2  \label{line1} \\
&=\int_{\delta/2}^{\infty}\t{d}\varphi \, \frac{1}{\delta}\int_{-\delta/2+4\alpha/L}^{\delta/2-4\alpha/L}\t{d} \eta  \, p_{\alpha,L}(\varphi-\eta)(\varphi+\delta/2)^2  \label{line2} \\
&=\frac{1}{\delta}\,\int_{\delta/2-\eta}^{\infty}\t{d}\varphi  \, p_{\alpha,L}(\varphi)\, \int_{-\delta/2+4\alpha/L}^{\delta/2-4\alpha/L}\t{d} \eta(\varphi+\eta+\delta/2)^2  \label{line3} \\
&=\frac{1}{\delta}\int_{4\alpha/L}^{\infty}\t{d}\varphi \, p_{\alpha,L}(\varphi)\int_{\max(-\delta/2+4\alpha/L,\delta/2-\varphi)}^{\delta/2-4\alpha/L}\t{d} \eta  \,  (\varphi+\eta+\delta/2)^2 \label{line4} \\
&=\frac{1}{\delta}\int_{4\alpha/L}^{\delta-4\alpha/L}\t{d}\varphi \, p_{\alpha,L}(\varphi)\int_{\delta/2-\varphi}^{\delta/2-4\alpha/L}\t{d} \eta  \,  (\varphi+\eta+\delta/2)^2 \nn
&\quad +\frac{1}{\delta}\int_{\delta-4\alpha/L}^{\infty} \t{d}\varphi \, p_{\alpha,L}(\varphi)\int_{-\delta/2+4\alpha/L}^{\delta/2-4\alpha/L}\t{d} \eta  \,  (\varphi+\eta+\delta/2)^2 \nn
&\le\frac{1}{\delta}\int_{4\alpha/L}^{\delta-4\alpha/L}\t{d}\varphi \, p_{\alpha,L}(\varphi) \frac 13 \left[\left( \varphi-4\alpha/L+\delta \right)^3-\delta^3\right] \nn
&\quad +\frac{1}{\delta}\int_{\delta-4\alpha/L}^{\infty} \t{d}\varphi \, p_{\alpha,L}(\varphi)\frac 13 \left[\left( \varphi-4\alpha/L+\delta \right)^3 -\left( \varphi+4\alpha/L \right)^3 \right].
\end{align}
In \eqref{line1} we have replaced all the estimated values with the extreme value $\tilde\varphi=-\delta/2$; in \eqref{line2} we have substituted the  exact formula for $p_{\alpha,L,\delta}(\varphi)$ \eqref{smeared}; in \eqref{line3} we have shifted the variable $\varphi\rightarrow\varphi+\eta$, and in \eqref{line4} we have modified the integration limits, while keeping the actual integration region unchanged.

\begin{figure*}[t!]
\includegraphics[width=0.5 \textwidth]{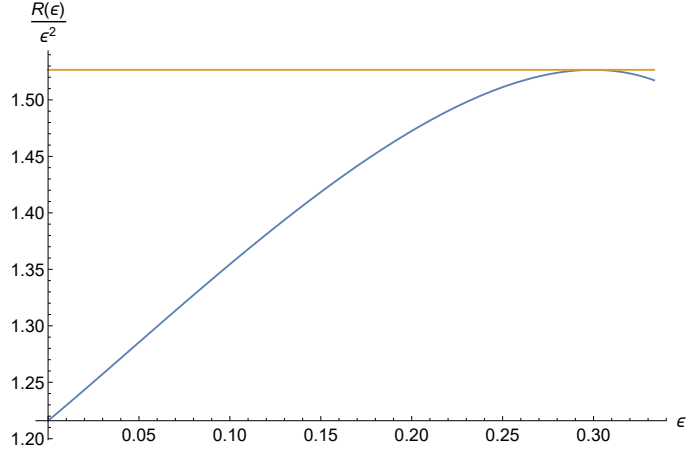}
\caption{\label{Repsilon}The ratio of $R(\epsilon)$ to $\epsilon^2$ for $\epsilon\in[0,1/3]$ (blue curve). The number $1.52661$ is shown as the horizontal orange line, showing that in this range $R(\epsilon)\leq 1.52661 \epsilon^2$.}
\end{figure*}

\begin{figure*}[t!]
\includegraphics[width=0.5 \textwidth]{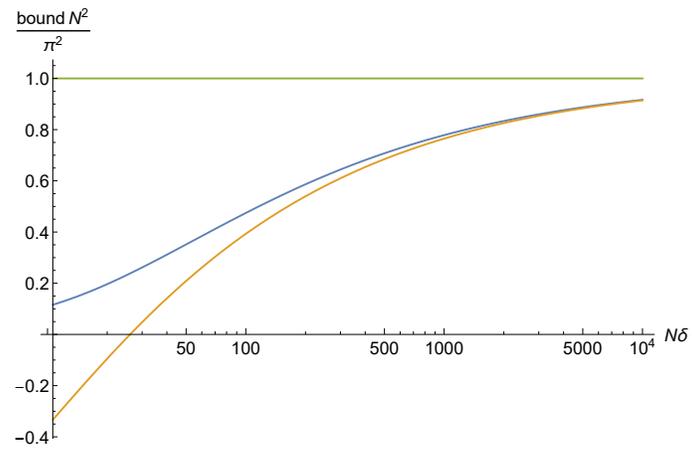}
\caption{\label{bound}This shows the comparison between the lower bound in Eq.~\eqref{bound1} (blue) and the lower bound in Eq.~\eqref{bound2} (orange), multiplied by $N^2/\pi^2$.
It can be seen that the lower bound in Eq.~\eqref{bound2} is lower than that in Eq.~\eqref{bound1}, so the additional $\widetilde{\mathcal{O}}$ term is not required.
The horizontal green line shows the asymptotic value of 1.}
\end{figure*}

Now, from the fact that $\forall_{\varphi\geq 4\alpha/L}p_{\alpha,L}(\varphi)=\mathcal N_\alpha L\,\sinc^4(\pi \alpha\sqrt{(L\varphi/4 \alpha)^2-1})$ and using inequalities $|\sinc(x)|\leq 1$ and $|\sinc(x)|\leq 1/|x|$, we may bound the above integral by
\begin{align}
&\frac{\mathcal N_\alpha L}{3\delta}\left(
\int_{4\alpha/L}^{8\alpha/L}\t{d}\varphi \,
 \left[\left( \varphi-4\alpha/L+\delta \right)^3- \delta^3\right]
 +  \int_{8\alpha/L}^{\delta-4\alpha/L}\t{d}\varphi \,  \frac{\left( \varphi-4\alpha/L+\delta \right)^3-\delta^3}{((L\varphi/4\alpha)^2-1)^2}\right. \nn & \qquad \left.
   +\int_{\delta-4\alpha/L}^{\infty} \t{d}\varphi \, \frac{\left( \varphi-4\alpha/L+\delta \right)^3 -\left( \varphi+4\alpha/L \right)^3 }{((L\varphi/4\alpha)^2-1)^2}\right).
 \end{align}
For notational clarity, let us introduce a small dimensionless parameter $\epsilon=\frac{4\alpha}{\delta L}$. The above expression now takes the form:
\begin{equation}
\label{integral}
\mathcal N_\alpha L\delta^3\underbrace{\frac{1}{3}\left(
\int_{\epsilon}^{2\epsilon}\t{d}t \,
 \left[\left( t-\epsilon+1 \right)^3-1\right]
+  \int_{2\epsilon}^{1-\epsilon}\t{d}t \,  \frac{\left(t-\epsilon+1 \right)^3-1}{((t/\epsilon)^2-1)^2}
  +\int_{1-\epsilon}^{\infty} \t{d}t \, \frac{\left( t-\epsilon+1 \right)^3 -\left(t+\epsilon \right)^3 }{((t/\epsilon)^2-1)^2}\right)}_{\frac{1}{2}R(\epsilon)}.
\end{equation}
Analogous reasoning holds for the left tail $\varphi\in[-\infty,-\delta/2]$. Therefore, we have:
\begin{equation}
\mathcal R_{\alpha,L,\delta}\leq \mathcal N_\alpha L\delta^3 R\left(\epsilon\right).
\end{equation}
The integral in \eqref{integral} may be calculated exactly and yields (assuming that $\epsilon\leq{1}/{3}$):
\begin{align}
R(\epsilon)=\frac{\epsilon^2}3 \left[3 + 5 \epsilon + \frac{7 \epsilon^2}6 + \left(\frac 1{2\epsilon} + 2 \epsilon^2 \right) \log(1 - 2 \epsilon) +
 \frac {3}2 \log 3  - 2 \epsilon^2\log(3 \epsilon) \right]
\end{align}
which may be bounded by $R(\epsilon)\leq 1.52661 \epsilon^2$ (see Figure \ref{Repsilon}).
After substituting into \eqref{boundcomp} and using $\mathcal N_\alpha \le {4\sqrt{2}\pi^4\alpha^{7/2}}{e^{-4\pi\alpha}}$ we have
\begin{equation}\label{bound1}
\Delta^2\varphi_\delta\geq \left(1-\frac{8\alpha}{\delta L}\right)\frac{\pi^2}{(n(\lambda_+-\lambda_-)+L/2)^2}-13460\frac{\delta \alpha^{11/2}}{L e^{4\pi\alpha}}\, ,
\end{equation}
which now may be optimized over $\alpha, L$ (with the restriction $\epsilon\leq 1/3$). In particular, defining $N=n(\lambda_+-\lambda_-)$, if we take
$L=\sqrt{8\alpha N/\delta}$, then
\begin{align}
 \Delta^2 \varphi_\delta &\ge \frac{\pi^2}{N^2}\left[ 1- 4\sqrt{\frac{2\alpha}{N\delta}}+\mathcal{O}\left(\alpha/N\delta\right)\right]-3365\sqrt 2 \frac{\delta^{3/2} \alpha^{5}}{N^{1/2} e^{4\pi\alpha}}\, .
\end{align}
If we then choose
\begin{equation}
\alpha = \frac 14 \log(N\delta),
\end{equation}
then the last term is higher order, and we get
\begin{align}
 \Delta^2 \varphi_\delta &\ge \frac{\pi^2}{N^2}\left[ 1- \sqrt{\frac{8\log(N\delta)}{N\delta}}+\widetilde{\mathcal{O}}\left((N\delta)^{-1}\right)\right] ,
\end{align}
using $\widetilde{\mathcal{O}}$ to indicate that logarithmic factors are omitted.
In fact, it is possible to give this inequality without the $\widetilde{\mathcal{O}}$ term as
\begin{align}\label{bound2}
 \Delta^2 \varphi_\delta &\ge \frac{\pi^2}{N^2
 }\left( 1- \sqrt{\frac{8\log(N\delta)}{N\delta}}\right) .
\end{align}
This inequality may be obtained numerically, by comparing the right-hand side (RHS) of Eq.~\eqref{bound1} to that of Eq.~\eqref{bound2} as a function of $N\delta$, as shown in Figure \ref{bound}.
This inequality is equivalent to Eq.~(14) of the main text.
The RHS of Eq.~\eqref{bound2} is only positive for $N\delta>26.0935$, so the inequality is only useful in that regime.
That also ensures that our assumption $\epsilon\le 1/3$ holds.
It is found that the RHS of Eq.~\eqref{bound1} is larger than that of Eq.~\eqref{bound2}, so Eq.~\eqref{bound1} implies Eq.~\eqref{bound2}.

\end{document}